# How to Combine Independent Data Sets for the Same Quantity

By

Theodore P. Hill[1] and Jack Miller[2]

**Abstract**

This paper describes a recent mathematical method called conflation for consolidating data from independent experiments that are designed to measure the same quantity, such as Planck's constant or the mass of the top quark. Conflation is easy to calculate and visualize, and minimizes the maximum loss in Shannon information in consolidating several independent distributions into a single distribution. In order to benefit the experimentalist with a much more transparent presentation than the previous mathematical treatise, the main basic properties of conflation are derived in the special case of normal (Gaussian) data. Included are examples of applications to real data from measurements of the fundamental physical constants and from measurements in high energy physics, and the conflation operation is generalized to weighted conflation for situations when the underlying experiments are not uniformly reliable.

1. School of Mathematics, Georgia Institute of Technology, Atlanta GA 303332.

2. Lawrence Berkeley National Laboratory, Berkeley CA 94720.

# 1. Introduction.

When different experiments are designed to measure the same unknown quantity, such as Planck's constant, how can their results be consolidated in an unbiased and optimal way? Given data from experiments that may differ in time, geographical location, methodology and even in underlying theory, is there a good method for combining the results from all the experiments into a single distribution?

Note that this is not the standard statistical problem of producing point estimates and confidence intervals, but rather simply to summarize all the experimental data with a single distribution. The consolidation of data from different sources can be particularly vexing in the determination of the values of the fundamental physical constants. For example, the U.S. National Institute of Standards and Technology (NIST) recently reported "two major inconsistencies" in some measured values of the molar volume of silicon $V_m(Si)$ and the silicon lattice spacing $d_{220}$, leading to an *ad hoc* factor of 1.5 increase in the uncertainty in the value of Planck's constant $h$ ([9, p. 54],[10]). (One of those two inconsistencies has subsequently been resolved [8].)

But input data distributions that happen to have different means and standard deviations are not necessarily "inconsistent" or "incoherent" [2, p 2249]. If the various input data are all normal (Gaussian) or exponential, for example, then every interval centered at the unknown positive true value has a positive probability of occurring in every one of the independent experiments. Ideally, of course, *all* experimental data, past as well as present, should be incorporated into the scientific record. But in the case of the fundamental physical constants, for instance, this could entail listing scores of past and present experimental datasets, each of which includes results from hundreds of experiments with thousands of data points, for *each one of the fundamental constants*. Most experimentalists and theoreticians who use Planck's constant, however, need a concise summary of its current value rather than the complete record. Having the mean and estimated standard deviation (e.g. via weighted least squares) does give some information, but without any knowledge of the distribution, knowing the mean within two standard deviations is only valid at the 75% level of significance, and knowing the mean within four standard deviations is not even significant at the standard 95% confidence level. Is there an objective, natural and optimal method for consolidating several input-data distributions into a single posterior distribution $P$? This article describes a new such method called *conflation*.

First, it is useful to review some of the shortcomings of standard methods for consolidating data from several different input distributions. For simplicity, consider the case of only two different experiments in which independent laboratories Lab I and Lab II measure the value of the same quantity. Lab I reports its results as a probability distribution $P_1$ (e.g. via an empirical histogram or probability density function), and Lab II reports its findings as $P_2$.

**Averaging the Probabilities**

One common method of consolidating two probability distributions is to simply average them - for every set of values *A*, set $P(A) = (P_1(A) + P_2(A))/2$. If the distributions both have densities, for example, averaging the probabilities results in a probability distribution with density the average of the two input densities (Figure 1). This method has several significant disadvantages. First, the mean of the resulting distribution *P* is always exactly the average of the means of $P_1$ and $P_2$, independent of the relative accuracies or variances of each. (Recall that the variance is the square of the standard deviation.) But if Lab I performed twice as many of the same type of trials as Lab II, the variance of $P_1$ would be half that of $P_2$, and it would be unreasonable to weight the two respective empirical means equally.

A second disadvantage of the method of averaging probabilities is that the variance of *P* is always *at least as large* as the minimum of the variances of $P_1$ and $P_2$ (see Figure 1), since $V(P) = [V(P_1) + V(P_2)]/2 + [mean(P_1) - mean(P_2)]^2/4$. If $P_1$ and $P_2$ are nearly identical, however, then their average is nearly identical to both inputs, whereas the standard deviation of a reasonable consolidation *P* should probably be strictly less than that of both $P_1$ and $P_2$. The method of averaging probabilities completely ignores the fact that two laboratories independently found nearly the same results. Figure 1 also shows another shortcoming of this method - with normally-distributed input data, it generally produces a multimodal distribution, whereas one might desire the consolidated output distribution to be of the same general form as that of the input data - normal, or at least unimodal.

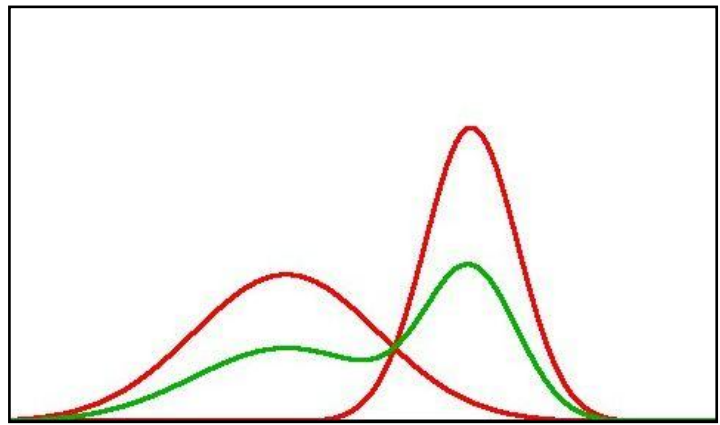

**Figure 1. Averaging the Probabilities.**

**(Green curve is the average of the red (input) curves. Note that the variance of the average is larger than the variance of either input.)**

**Averaging the Data**

Another common method of consolidating data - one that does preserve normality - is to average the underlying input data itself. That is, if the result of the experiment from Lab I is a random variable $X_1$ (i.e. has distribution $P_1$) and the result of Lab II is $X_2$ (independent of $X_1$, with distribution $P_2$), take $P$ to be the distribution of $(X_1 + X_2)/2$. As with averaging the distributions, averaging the data also results in a distribution that always has exactly the average of the means of the two input distributions, regardless of the relative accuracies of the two input data-set distributions (see Figure 2). With this method, on the other hand, the variance of $P$ is *never larger* than the maximum variance of $P_1$ and $P_2$ (since $V(P) = \left. V(P_1) + V(P_2) \right/ 4$), whereas some input data distributions that differ significantly should sometimes reflect a higher uncertainty. A more fundamental problem with this method is that in general it requires averaging data that was obtained using very different and even indirect methods, for example as with the watt balance and x-ray/optical interferometer measurements used in part to obtain the 2006 CODATA recommended value for Planck's constant.

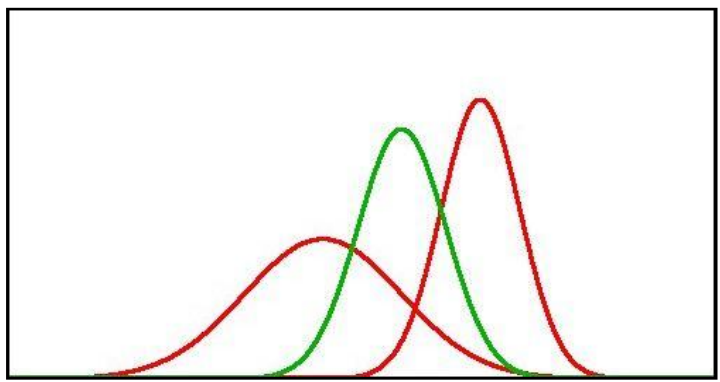

**Figure 2. Averaging the Data**

**(Green curve is the average of the red data. Note that the mean of the averaged data is exactly the average of the means of the two input distributions, even though they have different variances)**

The main goals of this paper are three: to describe conflation and derive important basic properties of conflation in the special case of normally-distributed data (perhaps the most common of experimental data); to provide concrete examples of conflation using real experimental data; and to introduce a new method for consolidating data when the underlying data sets are not uniformly weighted.

## 2. Conflation of Data Sets

For consolidating data from different independent sources, [6] introduced a mathematical method called *conflation* as an alternative to averaging the probabilities or averaging the data. Conflation (designated with the symbol "&" to suggest consolidation of $P_1$ and $P_2$) has none of the disadvantages of the two averaging methods described above, and has many advantages that will be described below.

In the important special case that the input distributions $P_1, P_2, ... P_n$ all have densities (e.g. normal or exponential distributions), then the conflation $\&(P_1, P_2, ..., P_n)$ of $P_1, P_2, ... P_n$ is simply the probability distribution with density the normalized product of the input densities. That is,

(*) If $P_1, ..., P_n$ have densities $f_1, ..., f_n$, respectively, and the denominator is not 0 or $\infty$, then

$$\&(P_1, P_2, ..., P_n) \text{ is continuous with density } f(x) = \frac{f_1(x)f_2(x) \cdots f_n(x)}{\int_{-\infty}^{\infty} f_1(y)f_2(y) \cdots f_n(y)dy}.$$

(Especially note that the product in (*) is taken for the densities evaluated at the same point, . Note also that conflation is easy to calculate, and to visualize; see Figure 3.)

**Remark.** For discrete input distributions, the analogous definition of conflation is the normalized product of the probability mass functions, and for more general situations the definition is more technical [6]. For the purposes of this paper, it will be assumed that the input distributions are continuous, and that the integral of their product is not 0 or $\infty$. This is always the case, for example, when the input distributions are all normal.

As can easily be seen from (*) and elementary conditional probability, the conflation of distributions has a natural heuristic and practical interpretation – gather data from the independent laboratories sequentially and simultaneously, and record the values only at those times when the laboratories (nearly) agree. This observation is readily apparent in the discrete case – if two independent integer-valued random variables $X_1$ and $X_2$ (e.g., binomial or Poisson random variables) have probability mass functions $f_1(k) = \Pr(X_1 = k)$ and $f_2(k) = \Pr(X_2 = k)$, then the probability that $X_1 = j$ given that $X_1 = X_2$, is simply $\dfrac{\Pr(X_1 = X_2 = j)}{\Pr(X_1 = X_2)} = \dfrac{f_1(j)f_2(j)}{\sum_k f_1(k)f_2(k)}$.

The argument in the continuous case follows similarly.

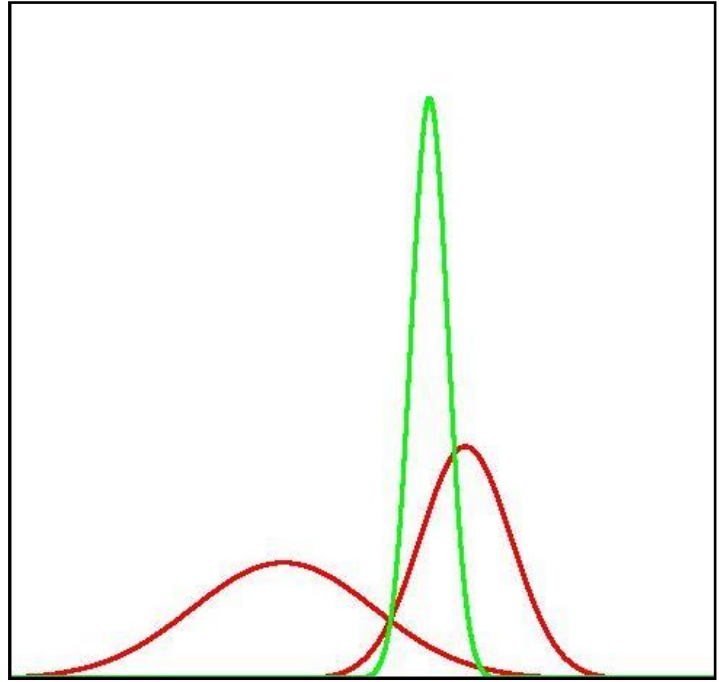

**Figure 3. Conflating Distributions**

**(Green curve is the conflation of red curves. Note that the mean of the conflation is closer to the mean of the input distribution with smaller variance, i.e. with greater accuracy)**

At first glance it may seem counterintuitive that the conflation of two relatively broad distributions can be a much narrower one (Figure 3). However, if both measurements are assumed equally valid, then with relatively high probability the true value should lie in the overlap region between the two distributions. Looking at it statistically, if one lab makes 50 measurements and another lab makes 100, then the standard deviations of their resulting distributions will usually be different. If the labs' methods are also different, with different systematic errors, or their methods rely on different fundamental constants with different uncertainties, then the means will likely be different too. But the bottom line is that the total of 150 valid measurements is substantially greater than either lab's data set, so the standard deviation should indeed be smaller.

## 3. Properties of Conflation

Conflation has several basic mathematical properties with significant practical advantages, and to describe these properties succinctly, it will be assumed throughout this section that $X_1$ and $X_2$ are independent normal random variables with means $m_1, m_2$ and standard deviations $\sigma_1, \sigma_2$, respectively. That is, for $i = 1, 2$, (**) $X_i \sim N(m_i, \sigma_i^2)$ has *density function* $f_i(x) = \dfrac{1}{\sigma_i\sqrt{2\pi}} \exp\left[\dfrac{(x-m_i)^2}{2\sigma_i^2}\right]$ for all $-\infty < x < \infty$

and *distribution* $P_i$ given by $P_i(A) = \int_A f_i(x)dx.$

**Remark.** The generalization of the properties of conflation described below to more than two distributions is routine; the generalization to non-normal distributions is more technical, and can be found in [6].

Some of the basic properties of conflation are as follows.

**(1) Conflation is commutative and associative:**

$$\&(P_1, P_2) = \&(P_2, P_1) \text{ and } \&(\&(P_1, P_2), P_3) = \&(P_1, \&(P_2, P_3)).$$

*Proof.* Immediate from (*) and the commutativity and associativity of real numbers, which implies that $f_1(x)f_2(x) = f_2(x)f_1(x)$ and $(f_1(x)f_2(x))f_3(x) = f_1(x)(f_2(x)f_3(x))$. □

**(2) Conflation is iterative:** $\&(P_1, P_2, P_3) = \&(\&(P_1, P_2), P_3) = \&(\&(\&(P_1), P_2), P_3).$

*Proof.* Immediate from (*). □

Thus from (2), to include a new data set in the consolidation, simply conflate it with the overall conflation of the previous data sets.

**(3) Conflations of normal distributions are normal:** If $P_1$ and $P_2$ satisfy (**), then

$\&(P_1, P_2)$ is normal with $m = \dfrac{\dfrac{m_1}{\sigma_1^2} + \dfrac{m_2}{\sigma_2^2}}{\dfrac{1}{\sigma_1^2} + \dfrac{1}{\sigma_2^2}} = \dfrac{\sigma_2^2 m_1 + \sigma_1^2 m_2}{\sigma_1^2 + \sigma_2^2}$ and $\sigma^2 = \dfrac{1}{\dfrac{1}{\sigma_1^2} + \dfrac{1}{\sigma_2^2}} = \dfrac{\sigma_1^2 \sigma_2^2}{\sigma_1^2 + \sigma_2^2}.$

*Proof.* By (*) and (**), $\&(P_1, P_2)$ is continuous with density proportional to

$f_1(x)f_2(x) = \dfrac{1}{\sigma_1\sigma_2 2\pi} \exp\left(-\left[\dfrac{(x-m_1)^2}{2\sigma_1^2}\right] - \left[\dfrac{(x-m_2)^2}{2\sigma_2^2}\right]\right).$ Completing the square of the exponent gives $-\left[\dfrac{(x-m_1)^2}{2\sigma_1^2}\right]-\left[\dfrac{(x-m_2)^2}{2\sigma_2^2}\right]=\left[-\left(\dfrac{1}{2\sigma_1^2}+\dfrac{1}{2\sigma_2^2}\right)x^2+\left(\dfrac{m_1}{2\sigma_1^2}+\dfrac{m_2}{2\sigma_2^2}\right)-\left(\dfrac{m_1^2}{2\sigma_1^2}+\dfrac{m_2^2}{2\sigma_2^2}\right)\right]=$

$\left[-\left(\dfrac{1}{2\sigma_1^2}+\dfrac{1}{2\sigma_2^2}\right)\left(x-\left(\dfrac{m_1}{\sigma_1^2}+\dfrac{m_2}{\sigma_2^2}\right)\Big/\left(\dfrac{1}{2\sigma_1^2}+\dfrac{1}{2\sigma_2^2}\right)\right)^2+\left(\dfrac{m_1}{\sigma_1^2}+\dfrac{m_2}{\sigma_2^2}\right)^2\Big/\left(\dfrac{1}{2\sigma_1^2}+\dfrac{1}{2\sigma_2^2}\right)-\left(\dfrac{m_1^2}{2\sigma_1^2}+\dfrac{m_2^2}{2\sigma_2^2}\right)\right],$

which is easily seen to be the exponent of the density of a normal distribution with the mean and variance in (3). □

By (2) and (3), conflations of any finite number of normal distributions are always normal (see Figure 3, and red curve in Figure 4B). Similarly, many of the other important classical families of distributions, including gamma, beta, uniform, exponential, Pareto, Laplace, Bernoulli, zeta and geometric families, are also preserved under conflation [7, Theorem 7.1].

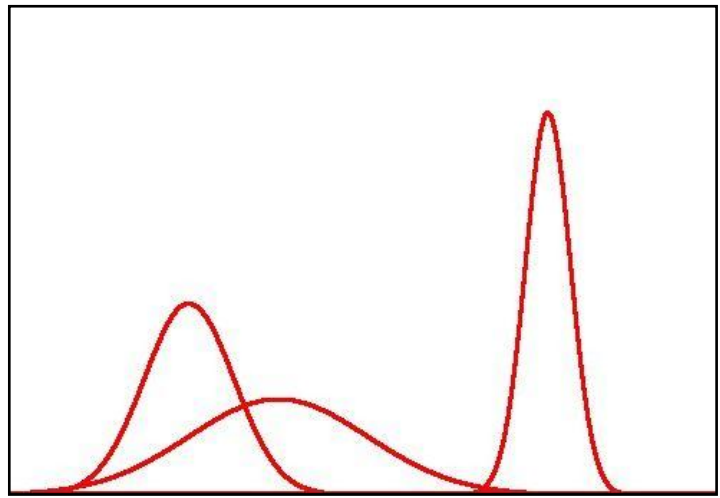

**Figure 4A. Three Input Distributions**

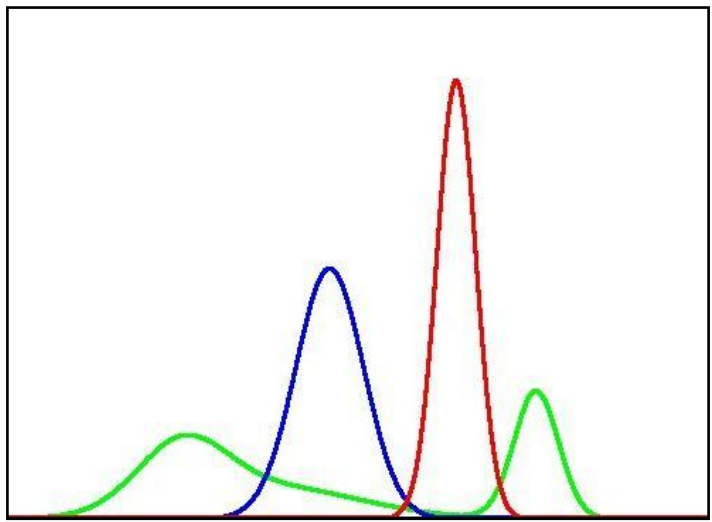

**Figure 4B. Comparison of Averaging Probabilities, Averaging Data, and Conflating**

**(Green curve is average of the three input distributions in Figure 4A, blue curve is average of the three input datasets, and red curve is the conflation.)**

**(4) Means and variances of conflations of normal distributions coincide with those of the weighted-least-squares method.**

*Sketch of proof.* Given two independent distributions with means $m_1, m_2$ and standard deviations $\sigma_1, \sigma_2$, respectively, the weighted-lease-squares mean $m$ is obtained by minimizing the function

$$f(m) = \frac{(m-m_1)^2}{\sigma_1^2} + \frac{(m-m_2)^2}{\sigma_2^2}$$ with respect to $m$. Setting $f'(m) = \frac{2(m-m_1)}{\sigma_1^2} + \frac{2(m-m_2)}{\sigma_2^2} = 0$

and solving for $m$ yields $m = \frac{\sigma_2^2 m_1 + \sigma_1^2 m_2}{\sigma_1^2 + \sigma_2^2}$, which, by (2), is the mean of the conflation of two normal distributions with means $m_1, m_2$ and standard deviations $\sigma_1, \sigma_2$. The conclusion for the weighted-least-squares variance follows similarly. □

**Shannon Information**

Whenever data from several (input) distributions is consolidated into a single (output) distribution, this will typically result in some loss of information, however that is defined. One of the most classical measures of information is the Shannon information. Recall that the Shannon information obtained from observing that a random variable *X* is in a certain set *A* is $-\log_2$ of the probability that *X* is in *A*. That is, the Shannon information is the number of binary bits of information obtained by observing that *X* is in *A*. For example, if *X* is a random variable uniformly distributed on the unit interval [0,1], then observing that *X* is greater than ½ has Shannon information exactly $-\log_2 \Pr(X > \tfrac{1}{2}) = -\log_2(\tfrac{1}{2}) = 1$, so one unit (binary bit) of Shannon information has been obtained, namely, that the first binary digit in the expansion of *X* is 1.

The Shannon Information is also called the *surprisal*, or *self-information* - the smaller the value of $\Pr(X \in A)$, the greater the information or surprise - and the (combined) Shannon Information obtained by observing that independent random variables $X_1$ and $X_2$ are *both* in *A* is simply the sum of the information obtained from each of the datasets $X_1$ and $X_2$, that is, $S_{P_1,P_2}(A) = S_{P_1}(A) + S_{P_2}(A) = -\log_2 P_1(A)P_2(A)$. Thus the *loss in Shannon information* incurred in replacing the pair of distributions $P_1, P_2$ by a single probability distribution $Q$ is $S_{P_1,P_2}(A) - Q(A)$ for the event *A*.

**(5) Conflation minimizes the loss of Shannon information:** *If $P_1$ and $P_2$ are independent probability distributions, then the conflation $\&(P_1, P_2)$ of $P_1$ and $P_2$ is the unique probability distribution that minimizes, over all events A, the maximum loss of Shannon information in replacing the pair $P_1, P_2$ by a single distribution Q.*

*Sketch of proof.* First observe that for an event *A*, the difference between the combined Shannon information obtained from $P_1$ and $P_2$ and the Shannon information obtained from a single probability *Q* is $S_{P_1,P_2}(A) - Q(A) = \log_2 \dfrac{Q(A)}{P_1(A)P_2(A)}$. Since $\log_2(x)$ is strictly increasing, the maximum (loss) thus occurs for an event *A* where $\dfrac{Q(A)}{P_1(A)P_2(A)}$ is maximized.

Next note that the largest loss of Shannon information occurs for small sets *A*, since for disjoint sets *A* and *B*,

$$\frac{Q(A \cup B)}{P_1(A \cup B)P_2(A \cup B)} \leq \frac{Q(A) + Q(B)}{P_1(A)P_2(A) + P_1(B)P_2(B)} \leq \max\left\{\frac{Q(A)}{P_1(A)P_2(A)}, \frac{Q(B)}{P_1(B)P_2(B)}\right\},$$

where the inequalities follow from the inequalities $(a+b)(c+d) \geq ac + bd$ and $\dfrac{a+b}{c+d} \leq \max\left\{\dfrac{a}{c}, \dfrac{b}{d}\right\}$ for positive numbers *a,b,c,d*. Since $P_1$ and $P_2$ are normal, their densities $f_1(x)$ and $f_2(x)$ are continuous everywhere, so the small set $A$ may in fact be replaced by an arbitrarily small interval, and the problem reduces to finding the probability density function $f$ that makes the maximum, over all real values $x$, of the ratio $\dfrac{f(x)}{f_1(x)f_2(x)}$ as small as possible. But, as is seen in the discrete framework, the minimum over all nonnegative $p_1,...,p_n$ with $p_1+...+p_n=1$ of the maximum of $\dfrac{p_1}{q_1},...,\dfrac{p_n}{q_n}$ occurs when $\dfrac{p_1}{q_1}=...=\dfrac{p_n}{q_n}$ (if they are not equal, reducing the numerator of the largest ratio, and increasing that of the smallest, will make the maximum smaller). Thus the $f$ that makes the maximum of $\dfrac{f(x)}{f_1(x)f_2(x)}$ as small as possible is when $f(x) = cf_1(x)f_2(x)$, where c is chosen to make $f$ a density function, i.e., to make $f$ integrate to 1. But this is exactly the definition of the conflation $\&(P_1,P_2)$ in (*). □

**Remark.** The proof only uses the facts that normal distributions have densities that are continuous and positive everywhere, and that the integral of the product of every two normal densities is finite and positive.

**(6) Conflation is a best linear unbiased estimate (BLUE):** *If $X_1$ and $X_2$ are independent unbiased estimates of $\theta$ with finite standard deviations $\sigma_1, \sigma_1$ respectively, then $\Theta = mean[\&(N_1,N_1)]$ is a best linear unbiased estimate for $\theta$, where $N_1$ and $N_2$ are independent normal probability distributions with (random) means $X_1$ and $X_2$, and standard deviations $\sigma_1$ and $\sigma_2$, respectively.*

*Sketch of proof.* Let $X = pX_1 + (1-p)X_2$ be the linear estimator of $\theta$ based on $X_1$ and $X_2$ and weight $0 \le p \le 1$. Then the expected value $E(X)$ of $X$ is $E(X) = pm_1 + (1-p)m_2$, and since $X_1$ and $X_2$ are independent the variance $V(X)$ of $X$ is $V(X) = p^2\sigma_1^2 + (1-p)^2\sigma_2^2$. To find the $p^*$ that minimizes $V(X)$, setting $\dfrac{dV}{dp} = 2p\sigma_1^2 - 2(1-p)\sigma_2^2 = 0$ yields $p = \dfrac{\sigma_2^2}{\sigma_1^2+\sigma_2^2}$, so $X^* = \dfrac{\sigma_2^2 X_1}{\sigma_1^2+\sigma_2^2} + \dfrac{\sigma_1^2 X_2}{\sigma_1^2+\sigma_2^2}$ is BLUE for $\theta$. But by (3), $X^*$ is the mean of $\&(N_1,N_2)$. □

**(7) Conflation yields a maximum likelihood estimator (MLE):** *If $X_1$ and $X_2$ are independent normal unbiased estimates of $\theta$ with finite standard deviations $\sigma_1, \sigma_1$ respectively, then $\Theta = mean[\&(N_1, N_1)]$ is a maximum likelihood estimator (MLE) for $\theta$, where $N_1$ and $N_2$ are independent normal probability distributions with (random) means $X_1$ and $X_2$, and standard deviations $\sigma_1$ and $\sigma_2$, respectively.*

*Sketch of proof.* The classical likelihood function in this case is

$$L = f(X_1;\theta)f(X_2;\theta) = \frac{1}{\sigma_1\sqrt{2\pi}}\exp\left[\frac{(X_1-\theta)^2}{2\sigma_1^2}\right]\frac{1}{\sigma_2\sqrt{2\pi}}\exp\left[\frac{(X_2-\theta)^2}{2\sigma_2^2}\right],$$

so to find the $\theta^*$ that maximizes $L$, take the partial derivative of $\log L$ with respect to $\theta$ and set it equal to zero -
$\frac{\partial \log L}{\partial \theta} = \frac{X_1-\theta}{\sigma_1^2} + \frac{X_2-\theta}{\sigma_2^2} = 0$. This implies that the critical point (and maximum likelihood) occurs when $\theta^*\left(\frac{1}{\sigma_1^2}+\frac{1}{\sigma_2^2}\right) = \left(\frac{X_1}{\sigma_1^2}+\frac{X_2}{\sigma_2^2}\right)$. Thus $\theta^* = \left(\frac{X_1}{\sigma_1^2}+\frac{X_2}{\sigma_2^2}\right) \bigg/ \left(\frac{1}{\sigma_1^2}+\frac{1}{\sigma_2^2}\right)$. By (3), this implies that the MLE $\theta^*$ is the mean of $\&(N_1, N_2)$. □

**Remark.** Note that the normality of the underlying distributions is used in (7), but it is not required for (5) or (6). Properties (4), (6), and (7) in the general cases use Aiken's generalization of the Gauss-Markov theorem and related results; e.g. see [1] and [11].

In addition to (6) and (7), conflation is also optimal with respect to several other statistical properties. In classical hypotheses testing, for example, a standard technique to decide from which of *n* known distributions given data actually came is to maximize the likelihood ratios, that is, the ratios of the probability density or probability mass functions. Analogously, when the objective is how best to consolidate data from those input distributions into a single (output) distribution $P$, one natural criterion is to choose $P$ so as to make the ratios of the likelihood of observing *x* under $P$ to the likelihood of observing *x* under *all* of the (independent) distributions $\{P_i\}$ as close as possible. The conflation of the distributions is the unique probability distribution that makes the variation of these likelihood ratios as small as possible.

The conflation of the distributions is also the unique probability distribution that preserves the proportionality of likelihoods. A criterion similar to likelihood ratios is to require that the output distribution *P* reflect the relative likelihoods of identical individual outcomes under the $\{P_i\}$. For example, if the likelihood of all the experiments $\{P_i\}$ observing the identical outcome *x* is twice that of the likelihood of all the experiments $\{P_i\}$ observing *y*, then *P(x)* should also be twice as large as *P(y)*.

Conflation has one more advantage over the methods of averaging probabilities or data. In practice, assumptions are often made about the form of the input distributions, such as an assumption that underlying data is normally distributed [9]. But the true and estimated values for Planck's constant are clearly never negative, so the underlying distribution is certainly not truly normally distributed – more likely it is truncated normal. Using conflations, the problem of truncation essentially disappears – it is automatically taken into account. If one of the input distributions is summarized as a true normal distribution, and the other excludes negative values, for example, then the conflation will exclude negative values, as is seen in Figure 5.

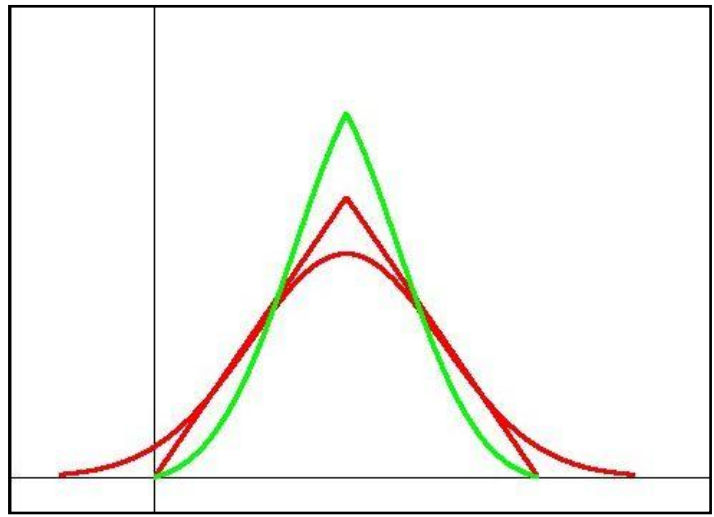

**Figure 5.** **(Green curve is the conflation of red curves. Note that the conflation has no negative values, since the triangular input had none.)**

## 4. Examples in Measurements of Physical Constants and High-energy Physics

As was described in the Introduction, methods for combining independent data sets are especially pertinent today as progress is made in creating highly precise measurement standards and reference values for basic physical quantites. The first two authors were originally concerned with the value of Avogadro's number [3] and later with a re-definition of the kilogram [7]. This endeavor brought them into contact with the researchers at NIST and their foreign counterparts, and, as suggested in the Introduction, it became apparent that an objective method for combining data sets measured in different laboratories is a pressing need.

While the authors of this paper believe that the kilogram should be defined in terms of a predetermined theoretical value for Avogadro's number [7], the NIST approach is based instead on a more precise value for Planck's constant determined in the laboratory using a watt balance. In fact this approach may result in a defined exact value for Planck's constant in parallel with the speed of light and the second (these two determine the meter exactly as well). Since *conflation* is the result produced by an objective analysis of exactly this question – how to consolidate data from independent experiments – perhaps conflation can be employed to obtain better consolidations of experimental data for the fundamental physical constants. The purpose of this section is to illustrate, using concrete real data, how conflation may be used for this problem.

**Example 1. {220} Lattice Spacing Measurements**

The input data used to obtain the CODATA 2006 recommended values and uncertainties of the fundamental physical constants includes the measurements and inferred values of the absolute {220} lattice spacing of various silicon crystals used in the determination of Planck's constant and the Avogadro constant. The four measurements came from three different laboratories, and had values 192,015.565(13), 192,015.5973(84), 192,015.5732(53) and 192,015.5685(67), respectively [10, Table XXIV], where the parenthetical entry is the uncertainty. The CODATA Task Force viewed the second value as "inconsistent" with the other three (see red curves in Figure 6) and made a consensus adjustment of the uncertainties. Since those values "are the means of tens of individual values, with each value being the average of about ten data points" [10], the central limit theorem suggests that the underlying datasets are approximately normally distributed as is shown in Figure 6 (red curves). The conflation of those four input distributions, however, requires no consensus adjustment, and yields a value essentially the same as the final CODATA value, namely, 192,015.5762 [10, Table LIII], but with a much smaller uncertainty. Since uncertainties play an important role in determining the *values* of the related constants via weighted least squares, this smaller, and theoretically justifiable, uncertainty is a potential improvement to the current accepted values.

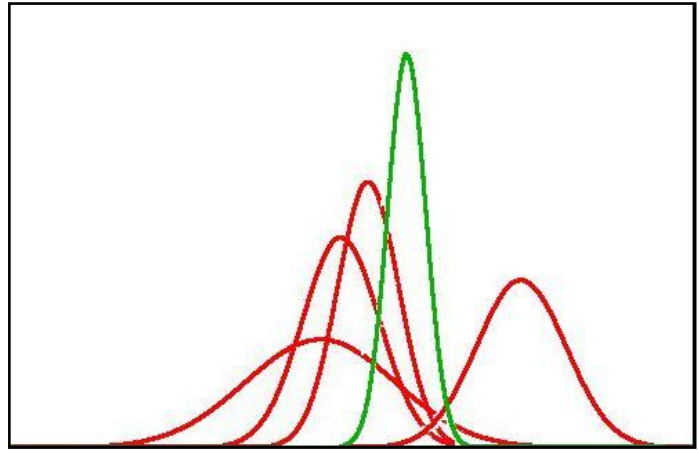

**Figure 6. (The four red curves are the distributions of the four measurements of the {220} lattice spacing underlying the CODATA 2006 values; the green curve is the conflation of those four distributions, and requires no ad hoc adjustment.)**

**Example 2. Top Quark Mass Measurements**

The top quark is a spin-1/2 fermion with charge two thirds that of the proton, and its mass is a fundamental parameter of the Standard Model of particle physics. Measurements of the mass of the top quark are done chiefly by two different detector groups at the Fermi national Accelerator laboratory (FNAL) Tevatron: the CDF collaboration using a multivariate-template method, a *b*-jet decay-length likelihood method, and a dynamic-likelihood method; and the DØ collaboration using a matrix-element-weighting method and a neutrino-weighting method. The mass of the top quark was then "calculated from eleven independent measurements made by the CDF and DØ collaborations" yielding the eleven measurements: 167.4(11.4), 168.4(12.8), 164.5(5.5), 178.1(8.3), 176.1(7.3), 180.1(5.3), 170.9(2.5), 170.3(4.4), 186.0(11.5), 174.0(5.2), and 183.9(15.8) GeV [6, Figure 4]. Again assuming that each of these measurements is approximately normally distributed, the conflation of these eleven independent input distributions is normal with mean and uncertainty (standard deviation) 172.63(1.6), which has a slightly higher mean and a lower uncertainty than the average mass of 171.4(2.1) reported in [5]. (Top quark measurements are being updated regularly, and the reader interested in the latest values should check the most recent FNAL publications; these concrete values were used simply for illustrative purposes.)

## 5. Weighted Conflation

The conflation $\&(P_1,...,P_n)$ of $n$ independent probability distributions (experimental datasets) $P_1,...,P_n$ described above and in [7] treated all the underlying distributions equally, with no differentiation between relative perceived validities of the experiments. A related statistical concept is that of a *uniform prior*, that is, a prior assumption that all the experiments are equally likely to be valid.

If, on the other hand, additional assumptions are made about the reliabilities or validities of the various experiments - for instance, that one experiment was supervised by a more experienced researcher, or employed a methodology thought to be better than another - then consolidating the data from the independent experiments should probably be adjusted to account for this perceived non-uniformity.

More concretely, suppose that in addition to the independent experimental distributions $P_1,...,P_n$, nonnegative weights $w_1,...,w_n$ are assigned to each of the distributions to reflect their perceived relative validity. For example, if $P_1$ is considered twice as reliable as $P_2$, then $w_1 = 2w_2$. Without loss of generality, the weights $w_1,...,w_n$ are nonnegative, and at least one is positive. How should this additional information $w_1,...,w_n$ be incorporated into the consolidation of the input data? That is, what probability distribution $Q = \&((P_1,w_1),...,(P_n,w_n))$ should replace the uniform-weight conflation $\&(P_1,...,P_n)$?

For the case where all the underlying datasets are assumed equally valid, it was seen that the conflation $\&(P_1,...,P_n)$ is the unique single probability distribution $Q$ that minimizes the loss of Shannon information between $Q$ and the original distributions $P_1,...,P_n$. Similarly, for weighted distributions $(P_1,w_1),...,(P_n,w_n)$, identifying the probability distribution $Q$ that minimizes the loss of Shannon information between $Q$ and the weighted data distributions leads to a unique distribution $\&((P_1,w_1),...,(P_n,w_n))$ called the *weighted conflation*.

Given $n$ weighted (independent) distributions $(P_1,w_1),...,(P_n,w_n)$, the *weighted Shannon Information* of the event $A$, $S_{((P_1,w_1),...,(P_n,w_n))}(A)$, is

$$S_{((P_1,w_1),...,(P_n,w_n))}(A) = \sum_{j=1}^{n} \frac{w_j}{w_{max}} S_{P_j}(A) = \sum_{j=1}^{n} -\left(\frac{w_j}{w_{max}}\right) \log_2 P_j(A),$$

where, here and throughout, $w_{max} = \max\{w_1,...,w_n\}$.

Note that $S_{((P_1,w_1),...,(P_n,w_n))}$ is continuous and symmetric in both $P_1,...,P_n$ and $w_1,...,w_n$, and that $S_{((P_1,w_1),...,(P_n,w_n))}(A) = 0$ if all the probabilities of $A$ are 1, for all $P_1,...,P_n$ and $w_1,...,w_n$. That is, no matter what the distributions and weights, no information is attained by observing any event that is certain to occur.

*Remarks*

(i) Dividing by $w_{max}$ reflects the assumption that only the *relative weights* are important, so for instance, if one experiment is considered twice as likely to be valid as another, then the information obtained from that experiment should be exactly twice as much as the information from the other, regardless of the absolute magnitudes of the weights. Thus in this latter case, for example, $S_{((P_1,2),(P_2,1))}(A) = S_{((P_1,4),(P_2,2))}(A) = S_{P_1}(A) + \frac{1}{2}S_{P_2}(A)$. In general, this means simply that for all $P_1,...,P_n$ and $w_1,...,w_n$,

$$S_{((P_1,w_1),...,(P_n,w_n))}(A) = S_{((P_1,w_1/w_{max}),...,(P_n,w_n/w_{max}))}(A).$$

(ii) If all the weights are equal, the weighted Shannon information coincides with the classical combined Shannon information, i.e.,

$$S_{((P_1,w_1),...,(P_n,w_n))}(A) = \sum_{j=1}^{n} S_{P_j}(A) \text{ if } w_1 = ... = w_n > 0.$$

(iii) The weighted Shannon information is at least the Shannon information of the single input distribution with the largest weight, and no more than the classical combined Shannon information of $P_1,...,P_n$, that is,

$$S_{P_1}(A) \leq S_{((P_1,w_1),...,(P_n,w_n))}(A) \leq S_{P_1,...,P_n}(A),$$

with equality if $w_1 > w_2 = ... = w_n = 0$, or $w_1 = ... = w_n > 0$, respectively.

Next, the basic definition of conflation (*) is generalized to the definition of weighted conflation, where $\&((P_1,w_1),...,(P_n,w_n))$ designates the *weighted conflation of* $P_1,...,P_n$ *with respect to the weights* $w_1,...,w_n$.

(***) If $P_1,...,P_n$ have densities $f_1,...,f_n$, respectively, and the denominator is not $0$ or $\infty$, then

$$\&((P_1,w_1),...,(P_n,w_n)) \text{ is continuous with density } f(x) = \frac{f_1^{\frac{w_1}{w_{\max}}}(x) f_2^{\frac{w_2}{w_{\max}}}(x) \cdots f_n^{\frac{w_n}{w_{\max}}}(x)}{\int_{-\infty}^{\infty} f_1^{\frac{w_1}{w_{\max}}}(y) f_2^{\frac{w_2}{w_{\max}}}(y) \cdots f_n^{\frac{w_n}{w_{\max}}}(y) dy}.$$

*Remarks.*

(i) The definition of weighted conflation for discrete distributions is analogous, with the probability density functions and integration replaced by probability mass functions and summation.

(ii) If $P_1,...,P_n$ are normal distributions with means $m_1,...,m_n$ and variances $\sigma_1^2,...,\sigma_n^2$ respectively, then an easy calculation shows that $\&((P_1,w_1),...,(P_n,w_n))$ is normal with

$$m = \frac{\frac{w_1 m_1}{\sigma_1^2} + ... + \frac{w_n m_n}{\sigma_n^2}}{\frac{w_1}{\sigma_1^2} + ... + \frac{w_n}{\sigma_n^2}} \quad \text{and} \quad \sigma^2 = \frac{w_{\max}}{\frac{w_1}{\sigma_1^2} + ... + \frac{w_n}{\sigma_n^2}}.$$

Observe that the mean of the weighted-conflation is closer to that of the mean of the distribution with the largest weight than the mean of the unweighted-conflation is, and the variance is also closer to the variance of that distribution. Also, an easy calculation shows that the variance of the weighted conflation is always at least as large as the variance of the equally-weighted conflation, and is never greater than the variance of the input distribution with the largest weight.

(iii) The weighted conflation depends only on the relative, not the absolute, values of the weights; that is

$$\&((P_1,w_1),...,(P_n,w_n)) = \&((P_1,w_1/w_{\max}),...,(P_n,w_n/w_{\max}))$$

(iv) If all the weights are equal, the weighted conflation coincides with the standard conflation, that is,

$$\&((P_1,w_1),...,(P_n,w_n)) = \&(P_1,...,P_n) \text{ if } w_1 = ... = w_n > 0.$$

(v) Updating a weighted distribution with an additional distribution and weight is straightforward: compute the weighted conflation of the pre-existing weighted conflation distribution and the new distribution, using weights $w_{\max} = \max\{w_1,...,w_n\}$ and $w_{n+1}$, respectively. That is, the analog of (2) for weighted conflation is

$$\&((P_1, w_1), \ldots, (P_n, w_n), (P_{n+1}, w_{n+1})) = \&(\&((P_1, w_1), \ldots, (P_n, w_n), w_{max}), (P_{n+1}, w_{n+1}))$$

(vi) Normalized products of density functions of the forms (*) and (***) have been studied in the context of "log opinion polls" and, more recently, in the setting of Hilbert spaces; see [4] and [7] and the references therein.

**(8) Weighted conflation minimizes the loss of weighted Shannon information:** *If $(P_1, w_1), \ldots, (P_n, w_n)$ are weighted independent distributions, then the weighted conflation $\&((P_1, w_1), \ldots, (P_n, w_n))$ is the unique probability distribution that minimizes, over all events A, the maximum loss of weighted Shannon information in replacing $(P_1, w_1), \ldots, (P_n, w_n)$ by a single distribution Q.*

The proofs of the above conclusions for weighted conflation follow almost exactly from those for uniform conflation, and the details are left for the interested reader.

## 5. Conclusion

The conflation of independent input-data distributions is a probability distribution that summarizes the data in an optimal and unbiased way. The input data may already be summarized, perhaps as a normal distribution with given mean and variance, or may be the raw data themselves in the form of an empirical histogram or density. The conflation of these input distributions is easy to calculate and visualize, and affords easy computation of sharp confidence intervals. Conflation is also easy to update, is the unique minimizer of loss of Shannon information, is the unique minimal likelihood ratio consolidation and is the unique proportional consolidation of the input distributions. Conflation of normal distributions is always normal, and conflation preserves truncation of data. Perhaps the method of *conflating* input data will provide a practical and simple, yet optimal and rigorous method to address the basic problem of consolidation of data.

### Acknowledgement

The authors are grateful to Dr. Peter Mohr for enlightening discussions regarding the 2006 CODATA evaluation process, and to Dr. Ron Fox for many suggestions.

Page **20** of **20**